\documentclass[a4paper]{article}

\usepackage{amsmath}
\usepackage{amsthm}
\usepackage{amsfonts} 
\usepackage{hyperref}
\usepackage{graphicx}
\usepackage{subcaption}
\usepackage{floatrow}
\usepackage{verbatimbox}
\usepackage{verbatim}
\usepackage{authblk}
\usepackage{booktabs}
\usepackage{verbments}

\newcommand\asteriskfill{\noindent\leavevmode\xleaders\hbox{$\ast$}\hfill\kern0pt}

\newcommand{\bftt}[1]{\textbf{\texttt{#1}}}

\usepackage[utf8]{inputenc}
\usepackage{pgfplots}
\DeclareUnicodeCharacter{2212}{−}
\usepgfplotslibrary{groupplots,dateplot}
\usetikzlibrary{patterns,shapes.arrows}
\pgfplotsset{compat=newest}

\title{Pabulib: A Participatory Budgeting Library\thanks{Please cite this paper when using Pabulib.}}

\author{%
  Dariusz Stolicki \\
  \small{Jagiellonian University in Kraków} \\
  \url{dariusz.stolicki@uj.edu.pl}
  \and
  Stanisław Szufa \\
  \small{Jagiellonian University in Kraków} \\
  \url{stanislaw.szufa@uj.edu.pl}
  \and
  Nimrod Talmon \\
  \small{Ben-Gurion University} \\
  \url{talmonn@bgu.ac.il}}

\begin{document}

\maketitle

\begin{abstract}
    We describe the \emph{PArticipatory BUdgeting LIBrary} website (in short, \emph{Pabulib}), which can be accessed via \url{http://pabulib.org/}, and which is a library of participatory budgeting data.
    In particular, we describe the file format (\texttt{.pb}) that is used for instances of participatory budgeting.
\end{abstract}

\section{Introduction}

Since it was initiated by the Brazil workers' party~\cite{wainwright2003making} in the 90s, Participatory budgeting (PB)~\cite{cabannes2004participatory} has been gaining increased attention all over the world.
Essentially, the idea behind PB is a direct democracy approach in which the way to utilize a common budget (most usually a municipality budget) is being decided upon by the stakeholders themselves (most usually city residents).
In particular, given a set of proposed projects with their costs, and a designated total budget to be used, voters express their preferences over the projects and then an aggregation method takes the votes and decides upon a subset of the projects to be implemented.

As research on PB from the perspective of computational social choice is accordingly increasing (see, e.g., the survey of Aziz and Shah~\cite{aziz2020participatory}; as well as some specific recent papers on PB~\cite{talmon2019framework,pbsub,goel2015knapsack,aziz2017proportionally}), there is a need to have publicly-available datasets;
this is the goal behind the \emph{PArticipatory BUdgeting LIBrary} (in short, \emph{Pabulib}), that is available in \url{http://pabulib.org}.

The main aim of this document is to define a data format that is used in Pabulib.

\section{The \texttt{.pb} File Format}

The data concerning one instance of participatory budgeting is to be stored in a single UTF-8 text file with the extension \texttt{.pb}.
The content of the file is to be divided into three sections:
\begin{itemize}
    \item \textbf{META} section with general  metadata like the country, budget, number of votes.
    \item \textbf{PROJECTS} section with projects costs and possibly some other metadata regarding projects like category, target etc.
    \item \textbf{VOTES} section with votes, that can be in one of the four types: approval, ordinal, cumulative, scoring; and optionally with metadata regarding voters like age, sex etc.
\end{itemize}

\section{A Simple Example}

\begin{Verbatim}[frame=single]
META 
key; value
description; Municipal PB in Wieliczka
country; Poland
unit; Wieliczka
instance; 2020
num_projects; 5
num_votes; 10
budget; 2500
rule; greedy
vote_type; approval
min_length; 1
max_length; 3
PROJECTS
project_id; cost; category 
1; 600; culture, education 
2; 800; sport
4; 1400; culture
5; 1000; health, sport
7; 1200; education
VOTES
voter_id; age; sex; vote
1; 34; f; 1,2,4
2; 51; m; 1,2
3; 23; m; 2,4,5
4; 19; f; 5,7
5; 62; f; 1,4,7
6; 54; m; 1,7
7; 49; m; 5
8; 27; f; 4
9; 39; f; 2,4,5
10; 44; m; 4,5
\end{Verbatim}

\section{Detailed Description}

The \textbf{bold} part is obligatory.

\subsection{Section 1: META}
    
    \begin{itemize}
        \item \bftt{key}
         \begin{itemize}
            \item \bftt{description}
            \item \bftt{country}
            \item \bftt{unit} -- name of the municipality, region, organization, etc., holding the PB process
            \item \texttt{subunit} -- name of the sub-jurisdiction or category within which the preferences are aggregated and funds are allocated
            \begin{itemize}
                \item \textit{Example}: in Paris, there are 21 PBs -- a city-wide budgets and 20 district-wide budgets. For the city-wide budget, \texttt{unit} is Paris, and \texttt{subunit} is undefined, while for the district-wide budgets, \texttt{unit} is also Paris, and \texttt{subunit} is the name of the district (e.g., IIIe arrondissement).
                \item \textit{Example}: before 2019, in Warsaw there have been district-wide and neighborhood-wide PBs. For all of them, \texttt{unit} is Warsaw, while \texttt{subunit} is the name of the district for district-wide budgets, and the name of the neighborhood for neighborhood-wide budgets. To associate neighborhoods with districts (if desired), an additional property \texttt{district} can be used.
                \item \textit{Example}: assume that in a given city, there are distinct PBs for each of $n>1$ categories (environmental projects, transportation projects, etc.). For all of them,  \texttt{unit} is the city name, while \texttt{subunit} is the name of the category.
            \end{itemize}
            \item \bftt{instance} -- a unique identifier of the specific edition of the PB process (year, edition number, etc.) used by the organizers to identify that edition; note that \texttt{instance} will not necessarily correspond to the year in which the vote is actually held, as some organizers identify the edition by the fiscal year in which the PB projects are to be carried out
            \item \bftt{num\_projects}
            \item \bftt{num\_votes}
            \item \bftt{budget} -- the total amount of funds to be allocated      
            \item \bftt{vote\_type}
            \begin{itemize}
                \item \texttt{approval} -- each vote is a vector of Boolean values, $\mathbf{v} \in \mathbb{B}^{|P|}$, where $P$ is the set of all projects,
                \item \texttt{ordinal} -- each vote is a permutation of a subset of $P$ such that $|P| \in [\mathtt{min\_length}, \mathtt{max\_length}]$, corresponding to a strict preference ordering,
                \item \texttt{cumulative} -- each vote is a vector $\mathbf{v} \in \mathbb{R}_{+}^{|P|}$ such that ${\lVert\mathbf{v}\rVert}_{1} \le \mathtt{max\_sum\_points} \in \mathbb{R}_{+}$,
                \item \texttt{scoring} -- each vote is a vector $\mathbf{v} \in I^{|P|}$, where $I \subseteq \mathbb{R}$.
            \end{itemize}
            \item \bftt{rule}  
            \begin{itemize}
                \item \texttt{greedy} -- projects are ordered decreasingly by the value of the aggregation function (i.e., the total score), and are funded until funds are exhausted or there are no more projects
                \item other rules will be defined in future versions
            \end{itemize}
            \item \texttt{date\_begin} -- the date on which voting starts
            \item \texttt{date\_end} -- the date on which voting ends
            \item \texttt{language} -- language of the description texts (i.e., full project names)
            \item \texttt{edition}
            \item \texttt{district}
            \item \texttt{comment}
            \item if \texttt{vote\_type} = \texttt{approval}:
                \begin{itemize}
                    \item \texttt{min\_length} [default: 1]
                    \item \texttt{max\_length} [default: num\_projects]
                    \item \texttt{min\_sum\_cost} [default: 0]
                    \item \texttt{max\_sum\_cost} [default: $\infty$]
                \end{itemize}
            \item if \texttt{vote\_type} = \texttt{ordinal}:
                \begin{itemize}
                    \item \texttt{min\_length} [default: 1]
                    \item \texttt{max\_length} [default: num\_projects]
                    \item \texttt{scoring\_fn} [default: Borda]
                \end{itemize}

            \item if \texttt{vote\_type} = \texttt{cumulative}:
                \begin{itemize}
                    \item \texttt{min\_length} [default: 1]
                    \item \texttt{max\_length} [default: num\_projects]
                    \item \texttt{min\_points} [default: 0]
                    \item \texttt{max\_points} [default: max\_sum\_points]
                    \item \texttt{min\_sum\_points} [default: 0]
                    \item \bftt{max\_sum\_points} 
                \end{itemize}
            \item if \texttt{vote\_type} = \texttt{scoring}:
                \begin{itemize}                    
                    \item \texttt{min\_length} [default: 1]
                    \item \texttt{max\_length} [default: num\_projects]
                    \item \texttt{min\_points} [default: $-\infty$]
                    \item \texttt{max\_points} [default: $\infty$]
                    \item \texttt{default\_score} [default: 0]
                \end{itemize}
            \item \texttt{non-standard fields}
        \end{itemize}
        \item \bftt{value}
    \end{itemize}

\subsection{Section 2: PROJECTS}
        
    \begin{itemize}
        \item \bftt{project\_id}
        \item \bftt{cost}
        \item \texttt{name} -- full project name
        \item \texttt{category} -- for example: education, sport, health, culture, environmental protection, public space, public transit and roads
        \item \texttt{target} -- for example: adults, seniors, children, youth, people with disabilities, families with children, animals
        \item \texttt{non-standard fields}
    \end{itemize}

\subsection{Section 3: VOTES}
    \begin{itemize}
        \item \bftt{voter\_id}
        \item \texttt{age}
        \item \texttt{sex}
        \item \texttt{voting\_method} (e.g., paper, Internet, mail)
        \item if \texttt{vote\_type} = \texttt{approval}:
            \begin{itemize}
                    \item \bftt{vote} -- ids of the approved projects, separated by commas.
            \end{itemize}
        \item if \texttt{vote\_type} = \texttt{ordinal}:
            \begin{itemize}
                    \item \bftt{vote} -- ids of the selected projects, from the most preferred one to the least preferred one, separated by commas.
            \end{itemize}
        \item if \texttt{vote\_type} = \texttt{cumulative}:
            \begin{itemize}
                \item \bftt{vote} -- project ids, in the decreasing order induced by \texttt{points}, separated by commas; projects not listed are assumed to be awarded $0$ points.
                \item \bftt{points} -- points assigned to the selected projects, listed in the same order as project ids in \bftt{vote}.
            \end{itemize}
        \item if \texttt{vote\_type} = \texttt{scoring}:
                    \begin{itemize}
                    \item \bftt{vote} -- project ids, in the decreasing order induced by \texttt{points}, separated by commas; projects not listed are assumed to be awarded \texttt{default\_score} points.
                    \item \bftt{points} -- points assigned to the selected projects, listed in the same order as project ids in \bftt{vote}.
            \end{itemize}
            
        \item \texttt{non-standard fields}
        
       \end{itemize}

\section{Outlook}

We have introduced the PArticipatory BUdgeting LIBrary (Pabulib; available at \url{http://pabulib.org}), and have described the \texttt{.pb} file format that is used in it.

We hope that Pabulib will foster meaningful research on PB, in particularly helping the computational social choice community offer better aggregation methods to be used in real-world instances of PB.

\section*{Acknowledgement}

Nimrod Talmon has been supported by the Israel Science Foundation (grant No. 630/19). Dariusz Stolicki and Stanis\l aw Szufa have been supported under the Polish Ministry of Science and Higher Education grant no. 0395/DLG/2018/10.

\bibliographystyle{plain}
\bibliography{bib}

\end{document}